# Google Classroom as a Tool of Support of Blended Learning for Geography Students

Olha V. Bondarenko[1][0000-0003-2356-2674], Svitlana V. Mantulenko[1][0000-0001-5673-0174] and Andrey V. Pikilnyak[2][0000-0003-0898-4756]

[1] Kryvyi Rih State Pedagogical University, 54, Gagarina Ave., Kryvyi Rih, 50086, Ukraine
bondarenko.olga@kdpu.edu.ua, mantulenkokdpu@ukr.net
[2] State Institution of Higher Education "Kryvyi Rih National University"
11, Vitali Matusevich St., Kryvyi Rih, 50027, Ukraine
pikilnyak@gmail.com

**Abstract.** The article reveals the experience of organizing blended learning for geography students using Google Classroom, and discloses its potential uses in the study of geography. For the last three years, the authors have tested such in-class and distance courses as "Cartography and Basics of Topography", "Population Geography", "Information Systems and Technologies in Tourism Industry", "Regional Economic and Social World Geography (Europe and the CIS)", "Regional Economic and Social World Geography (Africa, Latin America, Asia, Anglo-America, Australia and Oceania)", "Socio-Economic Cartography".

The advantages of using the specified interactive tool during the study of geographical disciplines are highlighted out in the article. As it has been established, the organization of the learning process using Google Classroom ensures the unity of in-class and out-of-class learning; it is designed to realize effective interaction of the subjects learning in real time; to monitor the quality of training and control the students' learning achievements in class as well as out of it, etc.

The article outlines the disadvantages that should be taken into account when organizing blended learning using Google Classroom, including the occasional predominance of students' external motivation in education and their low level of readiness for work in the classroom; insufficient level of material and technical support in some classrooms; need for out-of-class pedagogical support; lack of guidance on the content aspect of Google Classroom pages, etc.

Through the test series conducted during 2016-2017, an increase in the number of geography students with a sufficient level of academic achievements and a decrease of those with a low level of it was revealed.

## 1 Introduction

### 1.1 The Problem Statement

The topicality of the problem using the Google Classroom is determined by a wide range of problems that can be presented in the form of contradictions between the social requirements for geography students' professional training, its specific characteristics (the need for organizing systematic educational activities outside the specially equipped laboratories of higher educational institutions (HEI): field practices, integrated practices, etc.) and the prevalence of higher educational institutions providing traditional didactic forms, methods and tools; between the constant growth of the volume of students' independent and individual work and the need for the facilitation of all types of educational activities of geography students at any time and in any place of its course using the available and corresponding tools of information and communication technologies (ICT), including the mobile ones; between the potential use of modern ICTs and inadequate level of readiness for their implementation by university lecturers and students.

One of the ways to overcome these contradictions is the implementation of *combined geography training*, which, according to the research carried out by Andrii M. Struk [21], is understood as a Geography training technique, integrating the in-class and out-of-class educational activities, provided that a pedagogically balanced combination of traditional as well as innovative techniques for the in-class, distance and mobile training is carried out for the effective educational goals achievement.

### 1.2 Theoretical background

Some aspects of the problem under study are highlighted in the scientific articles devoted to theoretical and methodological principles and the methodology of distance education (Aleksandr A. Andreev [1], Myroslav I. Zhaldak [24], Volodymyr M. Kukharenko [3], Yukhym I. Mashbyts [8], Svitlana V. Shokaliuk [17]); blended training organization (Volodymyr M. Kukharenko [6], Natalia V. Rashevska [12], Serhii O. Semerikov [16], Andrii M. Struk [21], Yurii V. Tryus [23], Bohdan I. Shunevych [18]); development of information-and-education environment (Aleksandr A. Andreev [1], Kateryna I. Slovak [20], Mariia A. Kyslova [7], Liubov F. Panchenko [11], Maiia V. Popel [19], Mariia P. Shyshkina [19]); the use of innovative ICT in the educational process (Valerii Yu. Bykov [3], Illia O. Teplytskyi [14, 15]).

The interest in solving the chosen problem is caused by research on geography methods of teaching, which reveals tendencies of educational space renewal by means of informatization of higher education (Oleh M. Topuzov [22]); geographic information systems and technologies (Viktor M. Samoilenko [13]); possibilities of providing geography distance learning (Yurii A. Fedorenko [4]). The blended learning of geography is represented on a larger scale in the writings of foreign scholars: the implementation of combined learning in the study of geography in the first year (Phillipa Mitchell and Pip Forer [9]), the influence of combined geography teaching on critical thinking of students (Özgen Korkmaz and Ufuk Karakuş [5]) and others.

The analysis of scientific developments and information resources of domestic higher educational institutions suggests that Moodle is a traditional tool of supporting blended learning in higher education [10], although nowadays there are other alternative options for open learning management systems, Google Classroom in particular.

At the same time, Google Classroom as a tool of supporting the blended training for geography students, has not yet found a comprehensive study and full coverage in the scientific writings of domestic researchers.

### 1.3 The objective of the article

The objective of the article is to highlight the experience of supporting the blended training for geography students by using Google Classroom.

## 2 Presenting the Main Material

Google Classroom is an educational interactive tool that allows creating an informatively rich educational environment integrating the Google Docs text editor, Google Drive cloud storage, Gmail and other applications (YouTube, Google Sheets, Google Slides, Google Forms, etc.) [7].

In terms of the interactive on-line interaction the Google Classroom is to: ensure the integrity of classroom and out-of-class work (group, independent, individual, etc.); realize effective interaction of learning subjects in real time through: creating tasks for each particular course and group with hyperlink onto multimedia content; editing and commenting on the state of a student's tasks; compiling individual tasks into thematic modules; publishing announcements, questions, information digests, etc.; controlling the students' individual tasks in both classroom and out-of-class time; setting deadlines for each task; commenting on the revised multimedia content offered for the tasks; assessing students' academic achievements on a national or international scale; copying the academic achievements to the Google Sheets to generate statistical reports, visual monitoring of the quality of training [2].

Within the Google Classroom, the interaction of all learning subjects ("student – student", "student – student group", "teacher – student", "teacher – student group") takes place not only for distance education (training communication outside the HEI), but also for the traditional in-class learning (training communication within the HEI) using e-mail, electronic conferences and other Internet communication tools. The most common forms of learning tools provided by the Google Classroom include: e-mail (Google Mail), e-conferencing (Hangouts), Google Forms, communication via chats, and others.

The revealing of the content of geographic disciplines, as well as the monitoring and control of geography students' academic achievements, is possible through the implementation of computer-based learning tools, in particular content delivery tools (Docs, Drive, Presentations, YouTube), visualization tools (Presentations, Maps, YouTube, Earth), control tools (Sheets, Forms), etc. (Fig. 1).

According to the developed model there have been developed in-class and distance learning courses in the following geographic disciplines: "Cartography and Basics of Topography", "Population Geography", "Information Systems and Technologies in Tourism Industry", "Regional Economic and Social World Geography (countries of Europe and the CIS)", "Regional Economic and Social World Geography (Africa, Latin

America, Asia, Anglo-America, Australia and Oceania)", "Socio-Economic Cartography" which have been tested for three years of study.

Let's have a closer look at the specific features of implementing the blended teaching of geographic disciplines in Google Classroom.

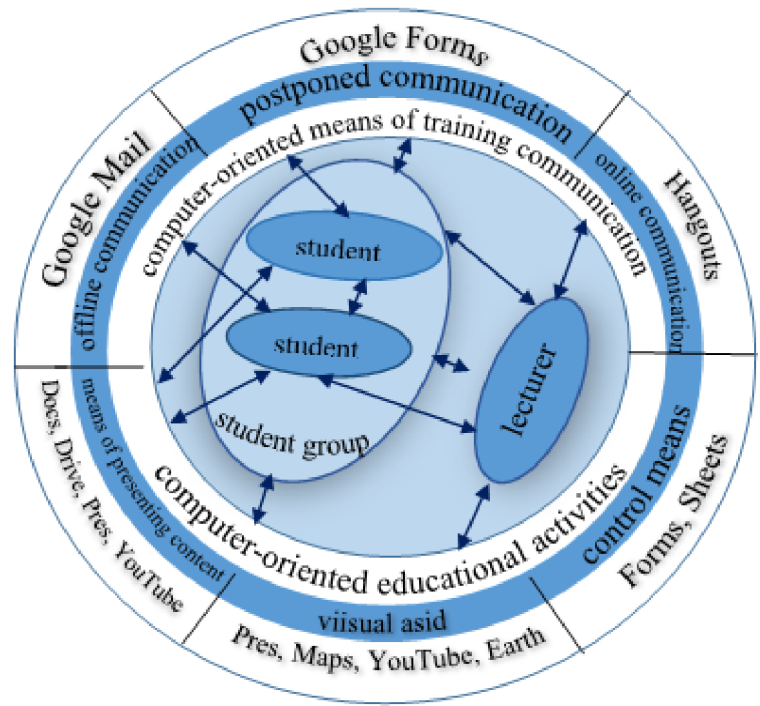

**Fig. 1.** The model of the information and education environment of blended learning for geography students in HEI based on Google Classroom

Each of the offered courses has a clearly defined structure. In the e-class, there are three pages "Stream", "Students", "Information", which have a certain content. So, the following basic elements are traditionally presented in "Stream": "practical / laboratory classes", "independent work", "nomenclature", "individual tasks" etc.

For every practical or laboratory lesson contained on the "Stream" page, you can not only add guidance to the tasks, but also attach any necessary file, vocation, and video, i.e., all the elements revealing the content of the subject being studied.

The "Students" page usually shows the class code and the contingent of the course attendees with access to their e-mail. The "Information" page, as a rule, provides such elements as "the course tasks and objectives, the classroom, the calendar / schedule of classes", "curriculum program", "list of recommended literature (basic, additional)", "contour maps", "maps and atlases", "reference sources", "methodical materials", "multimedia gallery", "Internet resources for creating maps". The contents of the page "Information" may vary depending on the specifics of the discipline content. The "Information" page contains electronic resources necessary for the tasks provided in "Stream" and provides a wide access to multimedia content, electronic libraries, textbooks, articles, maps, atlases, sites of international organizations, research institutes, databases, etc.

The extra benefits of using the specified resource for geographic courses is determined by the fact that most classes require work with contour maps, charts, diagrams, etc. In the Google Classroom, students can create maps by themselves using various editors and resources (DataGraf, Google Earth), tasks (learningapps.org); work with interactive maps (MigrationsMap, kartograph.org), statistical sources (USS\Ukrainian State Statistics Service, countrymeters.info); analytical data of international organizations (UNO, WHO, etc.); demonstrate knowledge of geographic nomenclature (online.seterra.net); conduct thematic control of knowledge (Google Forms) and others.

A compulsory element in geography students' professional training is the knowledge of the geographical nomenclature. As a rule, students pass the nomenclature by oral questioning using wall maps. The disadvantages of such a method of training are considered to be: a large time amount spent on the survey of one student and a group as a whole; the obsolete content of the wall social and economic maps; subjective assessment of knowledge of the nomenclature, etc.

Google Classroom allows replacing the traditional methodic of compiling the nomenclature for the interactive one. For example, second-year students are offered the Seterra online resource (https://online.seterra.com) and the Click-that-hood (http://click-that-hood.com) has been adjusted for the third-year students. The content of the task is that a student is to demonstrate the knowledge of the nomenclature within the time limit, save the version and send it to the teacher for marking.

The advantages of such a check of the nomenclature are: the individual pace of the task; objectivity of assessment; mapping skills; rational in-class time management.

When studying the above-mentioned disciplines, Google Classroom is used with a different didactic purpose. Thus, students of the first year use it with a propaedutic purpose, as a multimedia library (without downloading works and sending it for correcting analysis to the teacher). This is explained by the fact that in practical classes of the "Cartography and Basics of Topography" course geography students are, first of all, to be able to work with geographic maps and carry out topographical surveys of the area. Therefore, the Google Classroom use will in no way replace work with a map or field surveys. However, freshmen performing such tasks as the definition of the scale of distances and areas, orientation angles, absolute heights on a topographic map face various difficulties. Unfortunately, the degree of understanding of the new material in the classroom in the presence of pedagogical support is much higher than during the independent extra-curriculum knowledge acquisition. In addition, at home, the student is not able to work with most geodesic instruments, whereas in the practical class one needs to know not only their structure, but also use them in practice. The multimedia library content allows revising in full extend what has been learned in class, and it provides access to the video, which demonstrates the algorithm of topographic, promotes a better acquisition of means and methods of topographic survey of the area.

While studying the "Population Geography" course the main emphasis is placed on the fact that second-year students, unlike freshmen, should not only review the content of the study material and reproduce it, but also perform constructive tasks, find information in various sources characterizing the population of the world and particular countries, to analyze processes and to identify demographic tendencies, to characterize quantitative and qualitative indices of the country's population based on the aggregation of cartographic and statistical data – all of it is impossible without having access to relevant record-statistics, which are of a dynamic nature. However, independent search for the necessary data carried out by students often causes difficulties. The teacher can help: by restricting the search field by offering linking to the sites of reputable statistical organizations ("Information – Reference Sources" page).

While studying the "Regional Economic and Social Geography" and "Socio-economic Cartography" courses in Year III-IV for Bachelor Degree and Year I for Master Degree, geography students work with Google Classroom in a complex way: they perform the proposed tasks in the required editors (Docs, Sheets, Slides, etc.), send them to the teacher for checking up, comment on the multimedia content of the class, monitoring their academic achievements, offer discussion and data analysis of the information found during the self-search to the colleagues, etc. [2].

Google Classroom acquires a particular importance during the study of "Information Systems and Technologies in Tourism Industry" course, as it can be perceived from the course name, ICT is its inalienable part. Thus, within the Google Classroom it is convenient to consider the hardware and software of the automation work of tourist enterprises; to demonstrate the organizational and communicational provision of the work of the tourist office, etc. In studying this discipline, students learn to use office applications (Google Docs, Microsoft Office 365), specific products (Quick Sales 2.0, SELF-Agent, etc.), get acquainted with automated reservation systems in tourism (Amadeus, Galileo) and others.

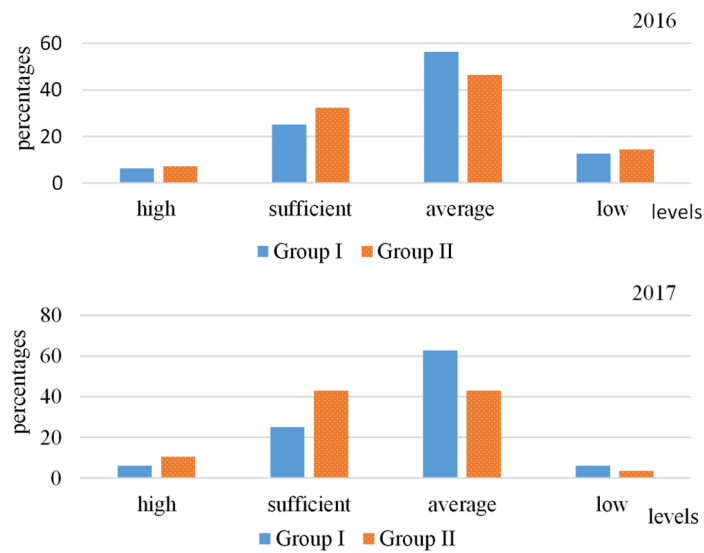

**Fig. 2.** Levels of students' academic achievements in the course "Regional Economic and Social World Geography"

At the final stage of the study of the mentioned above geographical disciplines during 2016 and 2017 the students' academic achievements have been monitored and summed up. As the proof of the developed distance learning courses effectiveness, the results of studying the course "Regional Economic and Social World Geography" are presented in Fig. 2. (traditional training method was used in Group I, and blended learning with Google Classroom – in Group II).

Analysis of Fig. 2 illustrates the positive dynamics in the levels of students' academic achievements in Group II. So, it is noticeable to observe an increase in the number of students with a sufficient level and a decrease in those who have shown a low level. In Group II, the number of high-level students increased from 7.1 % to 10.7 %; with sufficient – from 32.2 % to 42.9 %; with an average decreased from 46.4 % to 42.9 %; with a low – from 14.3 % to 3.5 %. There were no significant changes in Group I.

## 3  Conclusion

1. Summarizing the above stated, we may claim the benefits of using the Google Classroom for blended learning organization are as follows: real-time interaction of real-time learning subjects, which is particularly valuable if the volume of independent work is increased; the presence of constant pedagogical support and ensuring the integrity of both in-class and out-of-class work; increasing the visual aids in learning; development of critical thinking; formation of professional geographic competencies; attracting students to the familiar electronic environment with the use of ICT; operational control of educational achievements.
2. The disadvantages to take into account when organizing distance learning through the Google Classroom are to be considered: the predominance of external learning motivation and the low level of readiness of individual students for working in the new environment; lack of proper material and technical support for particular academic classrooms in HEI; the need for extra-curriculum pedagogical support, which requires additional time consuming from the teacher; inadequate attention of individual teachers to the problem of in-class and distance learning implementation.
3. Further study of the problem on organizing the blended learning for geography students is planned in the direction of developing a model and methodic of using Google Classroom as a tool of blended training future teachers of geography.